\documentclass[a4paper]{report}
\usepackage[utf8]{inputenc}
\usepackage[T1]{fontenc}
\usepackage{RJournal}
\usepackage{amsmath,amssymb,array}
\usepackage{booktabs}

\providecommand{\tightlist}{%
  \setlength{\itemsep}{0pt}\setlength{\parskip}{0pt}}

\usepackage{longtable}

\begin{document}

\sectionhead{Contributed research article}
\volume{XX}
\volnumber{YY}
\year{20ZZ}
\month{AAAA}

\begin{article}
\title{An Introduction to Rocker:\\
Docker Containers for R}
\author{by Carl Boettiger, Dirk Eddelbuettel}

\maketitle

\abstract{%
We describe the Rocker project, which provides a widely-used suite of
Docker images with customized R environments for particular tasks. We
discuss how this suite is organized, and how these tools can increase
portability, scaling, reproducibility, and convenience of R users and
developers.
}

\section{Introduction}\label{introduction}

The Rocker project was launched in October 2014 as a collaboration
between CB \& DE to provide high-quality Docker images containing the R
environment \citep{edd2014}. Since that time, the project has seen both
considerable uptake in the community and substantial development and
evolution. Here we seek to document the project's objectives and uses.

\subsubsection{What is Docker?}\label{what-is-docker}

Docker is a popular open-source tool to create, distribute, deploy, and
run software applications using \emph{containers}. Containers provide a
virtual environment (see \citet{Clark2014} for an overview of common
virtual environments) requiring all operating-system components an
application needs to run: code, runtime, system tools, system runtime.
Docker containers are lightweight as they share the operating system
kernel, starting instantly using a layered filesystem which minimizes
disk footprint and download time, are built on open standards that run
on all major platforms (Linux, Mac, Windows), and provide an added layer
of security by running an application in an isolated environment
\citep{what-docker}. Familiarity with a few key terms is helpful in
understanding this paper. The term ``container'' refers to an isolated
software environment on a computer. R users can think of running a
container as analogous to loading an R package; it makes a set of
software functions available. A Docker ``image'' is a binary archive of
that software, analogous to an R binary package: a given version is
downloaded only once, and can then be ``run'' to create a container
whenever it is needed. A ``Dockerfile'' is a recipe, the source-code, to
create a Docker image. Development and contributions to the Rocker
project focuses on the construction, organization and maintenance of
these Dockerfiles.

\section{Design principles \& use
cases}\label{design-principles-use-cases}

Docker gives users very convenient access to pre-configured and
pre-built \emph{binary} images that ``just work''. This allows R users
to access a wider-variety of ready-to-use environments than provided by
either the R Project itself or, say, their distribution which will
generally focus on one (current) release. For example, R users on
Windows may run RStudio Server or Shiny Server locally just by launching
a single command (once Docker itself is installed). Another common
use-case is access to R-devel without affecting the local system. Here,
we detail some of the principle use cases motivating these containerized
versions of R environments, and the design principles that help make
them work.

\subsubsection{Portability: From laptop to
cloud}\label{portability-from-laptop-to-cloud}

One common use case for Rocker containers is to provide a fast and
reliable mechanism to deploy a custom R environment to a remote server,
such as Amazon Web Services Elastic Compute (AWS EC2), DigitalOcean,
NSF's Jetstream servers \citep{jetstream}, or private or institutional
server hardware. Rocker containers are also easy to run locally on most
modern laptops by first installing the appropriate Docker command (or
distribution) for Windows, MacOS, and Linux-based operating systems. By
sharing volumes with the local host, users can still manipulate files
with familiar, native tools while performing computation through a
reproducible, containerized environment \citep{Boettiger2015}. Being
able to test code in a predictable, pre-configured R environment on a
local machine and to then run the same code in an identical environment
on a remote server (\emph{e.g.}, for access to greater RAM, more
processors, or merely to free up the local machine from a long-running
computation) is essential for low-friction scaling of analysis. Without
such containerization, getting code to run appropriately in a remote
environment can be a major undertaking, requiring both time and
knowledge many would-be users may not have.

For instance, the following bash commands install Docker on almost any
Linux-based server and then launch a Rocker container providing the
RStudio-server environment over a web interface.

\begin{verbatim}
wget -qO- https://get.docker.com/ | sh 
sudo docker run -p 8787:8787 -e PASSWORD=<PICK-A-PASSWORD> rocker/rstudio
\end{verbatim}

The \texttt{docker\ run} option \texttt{-p} sets the port on which
RStudio will appear, 8787 is the RStudio default (adding your user to
the docker group to avoids the need for a \texttt{sudo} command to call
\texttt{docker}: \texttt{sudo\ usermod\ -g\ docker\ \$USER}). Many
academic and commercial cloud providers make it possible to execute such
code snippets when a container is launched, without ever needing to
\texttt{ssh} into the machine. The user may log into the server merely
by pasting its IP address or DNS name (followed by the chosen port,
\emph{e.g.}, \texttt{:8787}) into a browser and entering the appropriate
password. This provides the user with a familiar, interactive
environment running on a remote machine while requiring a minimum of
expertise.

Deploying Docker containers on centrally administrated multi-user
machines, such as department or university clusters, has previously been
more difficult, since system adminstrators do not want to allow the
elevated user permissions the Docker runtime environment requires. To
work around this problem, Lawrence Berkeley National Labs (LBNL) has
made `Singularity' \citep{singularity}: a container runtime environment
that users can both install and use to run most Docker containers
without requiring root privileges, making it easier to deploy Rocker
containers in this context as well.

This portability is also valuable in an instructional context. Requiring
students to install all necessary software on personal laptops can be
particularly challenging for short workshops, where download and
installation time and troubleshooting across heterogeneous machines can
prove time consuming and frustrating for students and instructors alike.
By deploying a Rocker image or Rocker-derived image (see
\emph{Extensibility}) on a cloud machine, an instructor can easily
provide all students access to the pre-configured software environment
using only the browser on their laptops. This strategy has proven
effective in our own experience in both workshops and semester-length
courses. Similar Docker-based cloud deployments have been scaled to
courses of 100s of students, \emph{e.g.}, at Duke \citep{Mine} and UC
Berkeley \citep{data8}.

\subsubsection{Interfaces}\label{interfaces}

An important aspect of the Rocker project design is the ability for
users to interact with the software on the container through either an
interactive shell session (such as the R shell or a bash shell), or
through a web browser accessing the
RStudio\textsuperscript{\textregistered} Server integrated development
environment (IDE). Traditional remote and high-performance computing
workflows for R users have usually required the use of \texttt{ssh} and
a terminal-only interface, posing a challenge for interactive graphics
and a barrier to users unfamiliar with these tools and environments.
Accessing an RStudio\textsuperscript{\textregistered} container through
the browser removes these barriers. Rocker images include the
RStudio-server software pre-installed and configured with the explicit
permission of RStudio\textsuperscript{\textregistered} Inc.

Users can access a root \texttt{bash} shell in a running Rocker
container using

\begin{verbatim}
docker exec -ti <container-id> bash
\end{verbatim}

which can be useful for administrative tasks such as installing system
dependencies. All Rocker images can also be run as an interactive R,
RScript or bash shell without running RStudio, which can be useful for
batch jobs or for anyone who prefers that environment.

As with any interactive Docker container, users should specify the
interactive (\texttt{-i}) and terminal (\texttt{-t}, here combined with
interactive as \texttt{-ti}) flags, and specify the desired executable
environment (\emph{e.g.}, \texttt{R}, though other common options may be
\texttt{Rscript} or \texttt{bash}):

\begin{verbatim}
docker run --rm -ti rocker/tidyverse R
\end{verbatim}

This example shows the use of the \texttt{-\/-rm} flag to indicate that
the container should be removed when the interactive session is
finished. Details on sharing volumes, managing user permissions, and
more can be found on the Rocker website,
\url{https://rocker-project.org}.

\subsubsection{Sandboxed}\label{sandboxed}

Another feature of Rocker containers is the ability to provide a
sandboxed environment, isolated from software and potentially from other
data on the machine. Many users are reluctant to upgrade their suite of
installed packages, which may break their existing code or even their R
environment if the installation goes poorly. However, upgrading packages
and/or the R environment is often necessary to run analysis of a
colleague, or access more recent methods. Rocker offers an easy
solution. For instance, a user can run R code requiring the most recent
versions of R and related packages inside a Rocker container without
having to upgrade their local installations first. Conversely, one could
use Rocker to run code on an older R release with prior versions of R
packages, again without having to make any alteration to one's local R
install. Another common use case is to access a container with support
for particular options such as using gcc or clang compiler sanitizers
\citep{edd_sanitizers}. These requires R itself be built with
specialized settings that may not be not available or familiar to many R
users on their native system, but can be easily deployed by pulling the
Rocker images \texttt{rocker/r-devel-san} or
\texttt{rocker/r-devel-ubsan-clang}.

This sandboxing feature is also valuable in the remote computing
context, allowing system administrators to grant users freedom to
install software which requires root privileges inside a container,
while not granting them root access on the host machine. Root access is
required to launch Docker containers, though not to access containers
already running and providing some service such as RStudio. Users
logging into a container through the
RStudio\textsuperscript{\textregistered} interface do not by default
have root privileges, though are able to install R packages. Granting
these users root privileges in the container still leaves them sandboxed
from the host container. Further, Docker images can be deployed from a
machine without root privileges through the alternate container runtime
environment, Singularity \citep{singularity}. Unlike traditional virtual
machines, these containers do not impose a large footprint of reserved
resources as a typical host can easily support 100s of containers
\citep{what-docker}.

\subsubsection{Transparent}\label{transparent}

Users can easily determine the software stack installed on any Rocker
image by examining the associated Dockerfile recipe, which provides a
concise, human-readable record of the installation. All Rocker images
use automated builds through Docker Hub, which also acts as the central,
default repository distributing the images. Using automated builds
rather than uploading pre-built image binaries to the Docker Hub avoids
the potential for the build not to match the recipe. The corresponding
Dockerfile is visible both on the Docker Hub and in the linked GitHub
repository of the Rocker project (\url{https://github.com/rocker-org}),
which provides a transparent versioned history of all changes made to
these recipes, as well as documentation, a community wiki, and issue
trackers for discussing proposed changes, bugs, improvements to the
Dockerfiles and troubleshoot any issues users may encounter. Having
these public source files built automatically by a trusted provider
(Docker Hub), rather than built locally and uploaded only in binary form
provides is also useful from a security perspective in avoiding malware.

\subsubsection{Community Optimized}\label{community-optimized}

Having a shared, transparent computational environment created by a
publicly hosted, reproducible recipe facilitates community input into
configuration details. R and many of its packages and related software
can be configured with a wide range of options, compilers, different
linear-algebra libraries and so forth. While this flexibility reflects
varying needs, many users rely on default settings which may be
optimized more for simplicity of installation than performance. The
Rocker recipes reflect significant community input on these choices, as
well as the considerable experience and expertise of the R Debian
maintainer of 20 years in what configuration options to use. This helps
create a more finely tuned, optimized reference implementation of the R
environment as well as a platform for comparing and discussing these
concerns which are often overlooked elsewhere. Issues and Pull Requests
on the Rocker repositories on GitHub attest to these discussions and
improvements. Widespread use of the Rocker image helps promote both
testing of these choices and contributions further tweaking the
configuration from many members of the R community.

\subsubsection{Versioned}\label{versioned}

Access to specific versions of software can be important for users who
need computational reproducibility more than having the latest release
of any piece of software, since subsequent releases can alter the
behavior of code, introduce errors or otherwise alter previous results.
The versioned stack (\texttt{r-ver}, \texttt{rstudio},
\texttt{tidyverse}, \texttt{verse}, and \texttt{geospatial}) provides
images which are intended to build an identical software stack every
time, regardless of the release of new libraries and packages. Users
should specify an R version tag in the Docker image name to request a
version stable image, \emph{e.g.}, \texttt{rocker/verse:3.4.0}. If no
tag is explicitly requested, Docker will provide the image with the tag
\texttt{:latest}, which will always have the latest available versions
of the software (built nightly).

Users building on the version-tagged images will by default use the MRAN
snapshot mirror \citep{MRAN} associated with the most recent date for
which that image was current. This ensures that a Dockerfile building
\texttt{FROM\ rocker/verse:3.4.1} will only install R package versions
that were available on CRAN on 2017-06-30, that is, the day R 3.4.1 was
released. This default can of course be overwritten in the standard R
manner, \emph{e.g.}, by specifying a different CRAN mirror explicitly in
any command to \texttt{install.packages()}, or by adjusting the default
CRAN mirror in
\texttt{options(repo=\textless{}CRAN-MIRROR\textgreater{})} in an
\texttt{.Rprofile}. Note that the MRAN date associated with the current
release (\emph{e.g.}, \texttt{3.4.2} at the time of writing) will
continue to advance on the Docker-hub image until the next R release.
Software installed from \texttt{apt-get} in these images will come from
the the stable Debian release (\texttt{stretch} or \texttt{jessie}) and
thus not change versions (though it will receive security patches).
Packages installed from BioConductor using the \texttt{bioclite()}
utility will also install the version appropriate to the version of R
found on the system (the Bioconductor semi-annual release model avoids
the need for an MRAN mirror). Users installing packages from GitHub or
other sources can request a specific git release tag or hash for a more
reproducible build, or adopt an alternative approach such as
\texttt{packrat} \citep{packrat}. A more general discussion of the use
and limitations of Docker for computational reproducibility can be found
in \cite{Boettiger2015}.

\subsubsection{Extensible}\label{extensible}

Any portable computational environment faces an inevitable tension
between the ``kitchen sink problem'' at one extreme, and the ``discovery
problem'' on the other. A kitchen sink image seeks to accommodate too
many use cases in a single image. Such images are inevitably very large
and thus slow or difficult to deploy, maintain and optimize. At the
other extreme, providing too many specialized images makes it more
difficult for a user to discover the one they need. The Rocker project
seeks to avoid both of these problems by providing a carefully-curated
suite of images that an be easily extended by individuals and
communities.

To make extensions transparent and persistent, Rocker images can be
extended by any user by writing their own Dockerfiles based on an
appropriate Rocker image. The Dockerfiles in the Rocker stack should
themselves provide a simple of example of this. A user begins by
selecting an appropriate base image for their needs: if the
RStudio\textsuperscript{\textregistered} interface is desired, a user
might start with \texttt{FROM\ rocker/rstudio}; an image for testing a
particular C code in an R package might use
\texttt{FROM\ rocker/r-devel-san}, an image for reproducing a data
analysis will probably select a stable version tag in addition to an
appropriate base library, \emph{e.g.},:
\texttt{FROM\ rocker/tidyverse:3.4.1} Users can easily add additional
software to any running Rocker image using the standard R and Debian
mechanisms. Details on how to extend Rocker images can be found at
\url{https://rocker-project.org}.

Sharing these Dockerfiles can also facilitate the emergence of
extensions tuned to particular communities. For instance, the
\texttt{rocker/geospatial} image emerged from the input of a number of
Rocker users all adding common geospatial libraries and packages on top
of the existing Rocker images. This coalescence helped create a more
fine-tuned image with broad support for a wide range of commonly-used
data formats and libraries. Other community images are developed and
maintained independently of the Rocker project, such as the
\texttt{popgen} image of population-genetics-oriented software developed
by the National Evolutionary Synthesis Center (NESCent). Rocker images
are also being used as base Docker images in the NSF sponsored Whole
Tale project for reproducible computing \citep{wholetale}, and are
heavily used by the R-Hub project in automated package testing
\citep{rhub}.

\section{Rocker organization and
workflow}\label{rocker-organization-and-workflow}

The Rocker project consists of a suite of images built automatically by
and hosted on the Docker Hub, \url{https://hub.docker.com/r/rocker}.
Source Dockerfiles, supporting scripts and documentation are hosted on
GitHub under the organization \texttt{rocker-org},
\url{https://github.com/rocker-org}. The issue tracker and pull requests
are used for community input, discussions and contributions to these
images. The Rocker project wiki provides a place to synthesize
community-contributed documentation, use-cases, and other knowledge
about using the Rocker images.

\subsubsection{Images in the Rocker
Project}\label{images-in-the-rocker-project}

The Rocker project aims to provide a small core of Docker images that
serve as convenient `base' images on which other users can build custom
R environments by writing their own Dockerfiles, while also providing a
`batteries included' approach of images that can be used out of the box.
The challenges of balancing diverse needs driven by very different use
cases against the overarching goals of creating images that are still
sufficiently light-weight, easy to use and easy to maintain is a
difficult art. The implementation in both individual Rocker images and
image stacks can never perfect that balance for everyone, but today
reflects the considerable community input and testing over the past few
years.

All Rocker images are based on the Debian Linux distribution. The Debian
platform provides a small base image, the well-known \texttt{apt}
package management system and rich ecosystem of software libraries,
making it the base image of choice for Docker images, including many of
the ``official'' images maintained by Docker's own development team. The
Debian platform is also perhaps the best-supported Linux platform within
the R community, including an active \texttt{r-sig-debian} listserve.
The relatively long period between stable Debian releases (roughly two
years recently) means that software in the Debian stable (\emph{e.g.},
\texttt{debian:jessie}, \texttt{debian:stretch}) releases can lag
significantly behind current releases of popular software, including R.
More recent versions of packages can be found in the pre-release
distribution, \texttt{debian:testing}, while the very latest binary
builds can be found on \texttt{debian:unstable}. The Rocker project can
be largely divided into two stacks which address different needs,
reflected in which Debian distribution they are based on. The first
stack is based on \texttt{debian:testing}. The second, more
recently-introduced stack, is based only on Debian stable releases.
Rocker images always point to specific stable releases (\texttt{jessie},
\texttt{stretch}), and do not use the tag \texttt{debian:stable}, which
is a rolling tag that always points to the most recent stable version.
The different Rocker stacks have different aims and thus provide
different images, as shown in Tables 1 \& 2 below.

\subsubsection{\texorpdfstring{The \texttt{debian:testing}-based
images}{The debian:testing-based images}}\label{the-debiantesting-based-images}

The \texttt{debian:testing} stack aims to make the most efficient use of
upstream builds: the pre-compiled \texttt{.deb} binaries provided by the
Debian repositories. It is both quicker and easier to install software
from binaries, since the package manager (\texttt{apt}) manages the
necessary (binary) dependencies and bypasses the time-consuming process
of compiling from source. Basing this stack on \texttt{debian:testing}
means that much more recent versions of commonly-used libraries and
compilers are available as binaries than would be found in a Debian
stable release. In order to provide access to the most recent available
binaries, this stack uses apt-pinning \citep{apt_pinning} to allow the
\texttt{apt} package manager to also install binaries from
\texttt{debian:unstable}, which represents the most recent, bleeding
edge of packages built for Debian when necessary. For instance, the
\texttt{r-base} image provided by Rocker installs the most recent
version of R as a binary from Debian \texttt{unstable}. Similarly,
recent versions of many popular R packages can also be installed through
the package manager, \emph{e.g.}, \texttt{apt-get\ install\ r-cran-xml}.
This can be particularly helpful for packages with external system
dependencies (such as \texttt{libxml2-dev} in this example) which cannot
be installed from the R console as they are system dependencies rather
than R packages installed from within R. We should note, however, that
only about 500 of the over 11,000 CRAN packages are available as Debian
packages.

As the names \texttt{testing} and \texttt{unstable} imply, this approach
is not without challenges. The particular version of any given package
can change as packages move from \texttt{unstable} into
\texttt{testing}. New versions are sent to \texttt{unstable} during the
normal course of Debian development. This can occasionally break an
previously-working installation command in a Dockerfile until the
maintainer redirects the package manager to install a package from the
\texttt{unstable} sources that could previously be installed from
\texttt{testing}, or vice versa (using the \texttt{-t} option in
\texttt{apt}). That said, packages only migrate from \texttt{unstable}
to \texttt{testing} after a period of several days---and if the
migration and installation of the particular version is free of
interactions with other packages in their dependency graph. That way,
\texttt{unstable} serves as validation lab which leaves \texttt{testing}
reasonably stable yet current.

Relative to \texttt{stable}, the \texttt{testing} stack thus offers the
following advantages:

\begin{enumerate}
\def\labelenumi{\arabic{enumi}.}
\tightlist
\item
  These Dockerfiles are easy to develop and extend because almost all
  software can be installed through the package manager.
\item
  These Dockerfiles always install the most recent available software.
\item
  These Dockerfiles can almost always build relatively quickly.
\end{enumerate}

and these down-sides:

\begin{enumerate}
\def\labelenumi{\arabic{enumi}.}
\tightlist
\item
  These Dockerfiles require occasional minor maintenance to ensure
  successful builds. (\emph{e.g.}, changing an installation directive
  from \texttt{unstable} to \texttt{testing} or vice versa).
\item
  The resulting images are inherently dynamic: rebuilding the same
  Dockerfile months or years apart will generate images with
  significantly different versions of software installed.
\item
  The use of apt pinning may be unfamiliar to some users, where the user
  must ultimately be responsible to ensure compatibility
  \citep{apt_pinning}.
\end{enumerate}

\subsubsection{Images overview}\label{images-overview}

The \texttt{debian:testing}-based stack currently includes seven images
actively maintained by the Rocker development team (Table 1).
\texttt{r-base} builds on \texttt{debian:testing}, and the other six in
the stack each build directly from \texttt{r-base}. The \texttt{r-base}
image is unique in that it is designated as the official image for the R
language by the Docker organization itself. This official image is
reviewed and then built by employees of Docker Inc based on a Dockerfile
maintained by the Rocker team. Consequently, users should refer to this
image in Docker commands without an organization namespace, \emph{e.g.},
\texttt{docker\ run\ -t\ r-base} to access the official image. All other
images in the Rocker project are not individually reviewed and built by
Docker Inc and must be referenced using the \texttt{rocker} namespace,
\emph{e.g.}, \texttt{docker\ run\ -ti\ rocker/r-devel}.

Several of the images in this stack are oriented towards the R
development community: \texttt{r-devel}, \texttt{drd},
\texttt{r-devel-san} \& \texttt{r-devel-ubsan-clang} all add a copy of
the development version of R side-by-side the current release of R
provided by \texttt{r-base}. On these images, the development version is
aliased to \texttt{RD} to distinguish from the current release,
\texttt{R}. As the names suggest, each provide slightly different
configurations. Of particular interest are the images providing
development R built with support for C/C++ address and
undefined-behavior sanitizers, which are somewhat more difficult to
configure \citep{edd_sanitizers}.

As these images focus on developers and/or as base images for custom
uses, this stack does not include many specific R packages. Additional
dependencies and packages can easily be installed from \texttt{apt}. R
packages not available in the \texttt{apt} repositories can be installed
directly from CRAN using either \texttt{R} or the \texttt{littler}
scripts, as described in \url{https://rocker-project.org/use}.

This stack also includes the images \texttt{shiny} and
\texttt{rstudio:testing} that provide Shiny server and
RStudio\textsuperscript{\textregistered} server IDE from
RStudio\textsuperscript{\textregistered} Inc, built on the
\texttt{r-base} image. RStudio\textsuperscript{\textregistered} and
Shiny are registered trademarks of RStudio Inc, and their use and the
distribution of their software in binary form on Docker Hub has been
granted to the Rocker project by explicit permission from RStudio. Users
should review RStudio's trademark use policy
(\url{http://www.rstudio.com/about/trademark/}) and address inquiries
about further distribution or other questions to
\href{mailto:permissions@rstudio.com}{\nolinkurl{permissions@rstudio.com}}.
The Rocker project also provides images with
RStudio\textsuperscript{\textregistered} server and Shiny server in the
stable versioned stack.

\textbf{Build schedule}: The official \texttt{r-base} image is rebuilt
by Docker following any updates to the official \texttt{debian} images
(roughly every few weeks). The rest of the stack uses build triggers
that rebuild the images either whenever \texttt{r-base} is updated, or
the Dockerfile sources are updated on the corresponding GitHub
repository. The only exception in this stack is the \texttt{drd} image,
which is rebuilt each week by a \texttt{cron} trigger.

\begin{longtable}[]{@{}llll@{}}
\caption{The \texttt{debian:testing} image stack}\tabularnewline
\toprule
\begin{minipage}[b]{0.20\columnwidth}\raggedright\strut
image\strut
\end{minipage} & \begin{minipage}[b]{0.50\columnwidth}\raggedright\strut
description\strut
\end{minipage} & \begin{minipage}[b]{0.08\columnwidth}\raggedright\strut
size\strut
\end{minipage} & \begin{minipage}[b]{0.10\columnwidth}\raggedright\strut
downloads\strut
\end{minipage}\tabularnewline
\midrule
\endfirsthead
\toprule
\begin{minipage}[b]{0.20\columnwidth}\raggedright\strut
image\strut
\end{minipage} & \begin{minipage}[b]{0.50\columnwidth}\raggedright\strut
description\strut
\end{minipage} & \begin{minipage}[b]{0.08\columnwidth}\raggedright\strut
size\strut
\end{minipage} & \begin{minipage}[b]{0.10\columnwidth}\raggedright\strut
downloads\strut
\end{minipage}\tabularnewline
\midrule
\endhead
\begin{minipage}[t]{0.20\columnwidth}\raggedright\strut
\href{https://hub.docker.com/r/_/r-base}{r-base}\strut
\end{minipage} & \begin{minipage}[t]{0.50\columnwidth}\raggedright\strut
official image with current version of R\strut
\end{minipage} & \begin{minipage}[t]{0.08\columnwidth}\raggedright\strut
254 MB\strut
\end{minipage} & \begin{minipage}[t]{0.10\columnwidth}\raggedright\strut
632,000\strut
\end{minipage}\tabularnewline
\begin{minipage}[t]{0.20\columnwidth}\raggedright\strut
\href{https://hub.docker.com/r/rocker/r-devel}{r-devel}\strut
\end{minipage} & \begin{minipage}[t]{0.50\columnwidth}\raggedright\strut
R-devel added side-by-side r-base (using alias \texttt{RD})\strut
\end{minipage} & \begin{minipage}[t]{0.08\columnwidth}\raggedright\strut
1 GB\strut
\end{minipage} & \begin{minipage}[t]{0.10\columnwidth}\raggedright\strut
4,000\strut
\end{minipage}\tabularnewline
\begin{minipage}[t]{0.20\columnwidth}\raggedright\strut
\href{https://hub.docker.com/r/rocker/drd}{drd}\strut
\end{minipage} & \begin{minipage}[t]{0.50\columnwidth}\raggedright\strut
lightweight r-devel, built weekly\strut
\end{minipage} & \begin{minipage}[t]{0.08\columnwidth}\raggedright\strut
571 MB\strut
\end{minipage} & \begin{minipage}[t]{0.10\columnwidth}\raggedright\strut
4,000\strut
\end{minipage}\tabularnewline
\begin{minipage}[t]{0.20\columnwidth}\raggedright\strut
\href{https://hub.docker.com/r/rocker/r-devel-san}{r-devel-san}\strut
\end{minipage} & \begin{minipage}[t]{0.50\columnwidth}\raggedright\strut
as r-devel, but built with compiler sanitizers\strut
\end{minipage} & \begin{minipage}[t]{0.08\columnwidth}\raggedright\strut
1.1 GB\strut
\end{minipage} & \begin{minipage}[t]{0.10\columnwidth}\raggedright\strut
1,000\strut
\end{minipage}\tabularnewline
\begin{minipage}[t]{0.20\columnwidth}\raggedright\strut
\href{https://hub.docker.com/r/rocker/r-devel-ubsan-clang}{r-devel-ubsan-clang}\strut
\end{minipage} & \begin{minipage}[t]{0.50\columnwidth}\raggedright\strut
Sanitizers, clang c compiler (instead of gcc)\strut
\end{minipage} & \begin{minipage}[t]{0.08\columnwidth}\raggedright\strut
1.1 GB\strut
\end{minipage} & \begin{minipage}[t]{0.10\columnwidth}\raggedright\strut
525\strut
\end{minipage}\tabularnewline
\begin{minipage}[t]{0.20\columnwidth}\raggedright\strut
\href{https://hub.docker.com/r/rocker/r-devel-san}{rstudio:testing}\strut
\end{minipage} & \begin{minipage}[t]{0.50\columnwidth}\raggedright\strut
rstudio on debian:testing\strut
\end{minipage} & \begin{minipage}[t]{0.08\columnwidth}\raggedright\strut
1.1 GB\strut
\end{minipage} & \begin{minipage}[t]{0.10\columnwidth}\raggedright\strut
1,000\strut
\end{minipage}\tabularnewline
\begin{minipage}[t]{0.20\columnwidth}\raggedright\strut
\href{https://hub.docker.com/r/rocker/shiny}{shiny}\strut
\end{minipage} & \begin{minipage}[t]{0.50\columnwidth}\raggedright\strut
shiny-server on r-base\strut
\end{minipage} & \begin{minipage}[t]{0.08\columnwidth}\raggedright\strut
409 MB\strut
\end{minipage} & \begin{minipage}[t]{0.10\columnwidth}\raggedright\strut
123,000\strut
\end{minipage}\tabularnewline
\bottomrule
\end{longtable}

\begin{longtable}[]{@{}llll@{}}
\caption{The \texttt{rocker-versioned} stack of images}\tabularnewline
\toprule
image & description & size & downloads\tabularnewline
\midrule
\endfirsthead
\toprule
image & description & size & downloads\tabularnewline
\midrule
\endhead
\href{https://hub.docker.com/r/rocker/r-ver}{r-ver} & Version-stable
base R \& src build tools & 219 MB & 6,000\tabularnewline
\href{https://hub.docker.com/r/rocker/rstudio}{rstudio} & Adds rstudio &
334 MB & 314,000\tabularnewline
\href{https://hub.docker.com/r/rocker/tidyverse}{tidyverse} & Adds
tidyverse \& devtools & 656 MB & 83,000 \footnote{This figure includes
  49,000 downloads under the earlier name \texttt{hadleyverse}.}\tabularnewline
\href{https://hub.docker.com/r/rocker/verse}{verse} & Adds java, tex \&
publishing-related packages & 947 MB & 9,000\tabularnewline
\href{https://hub.docker.com/r/rocker/geospatial}{geospatial} & Adds
geospatial libraries & 1.3 GB & 4,000\tabularnewline
\bottomrule
\end{longtable}

\subsubsection{\texorpdfstring{The \texttt{debian:stable}-based
stack}{The debian:stable-based stack}}\label{the-debianstable-based-stack}

This stack emphasizes stability and reproducibility of the Docker build.
This stack was introduced much more recently (November 2016) in response
to considerable user input and requests. The key feature of this stack
is the ability to run older versions of R along with the
then-contemporaneous versions of R packages. A user specifies the
version desired using an image tag, \emph{e.g.},
\texttt{rocker/r-ver:3.3.1} will refer to an image with R version 3.3.1
installed. Omitting the tag is equivalent to using the tag
\texttt{latest}, which, as the name implies, will always point to an
image using the current R release. Thus, users wanting to create
downstream Dockerfiles which are based on the current release at time
(but will continue to reconstruct the same environment in the future
after newer R versions are released) should explicitly include the
corresponding version tag, \emph{e.g.}, \texttt{rocker/r-ver:3.4.2} at
the time of writing, and not the \texttt{latest} tag. Users can also run
the current development version of R using the tag \texttt{devel}, which
is built nightly from R-devel sources from \texttt{subversion}.

\textbf{MRAN archives}: To facilitate installation of only
contemporaneous versions of R packages on these images, the default CRAN
mirror from which to install R packages is fixed to a snapshot of CRAN
corresponding to the last date for which that version of R was the
latest release. These snapshots are provided by the MRAN archive created
by Revolution Analytics (now part of Microsoft). It archives daily
snapshots of all of CRAN from which a user can install packages with the
usual \texttt{install.packages()} function \citep{MRAN}. Users can
always override this default by passing any current CRAN repository
explicitly. Unlike CRAN, Bioconductor only updates its repositories
through bi-annual releases aligned to R's spring release schedule. Thus,
Bioconductor packages can be installed in the usual way using
\texttt{bioclite}, which automatically selects the Bioconductor release
corresponding to the version of R in use.

\textbf{Version tags}: The version tags are propagated throughout this
stack: \emph{e.g.}, \texttt{rocker/tidyverse:devel} will provide the
currently-released versions of the R packages in the \texttt{tidyverse}
\citep{tidyverse} installed on the nightly build of R-devel. Developers
building packages on this stack are encouraged to tag their images
accordingly as well. Table 3 indicates which versions of R are currently
available in the stack, going back to \texttt{3.1.0}. While older
versions may be added to the stack at a later date, we note that the
MRAN snapshots began in 2014-09-17 and thus go back only to the R
\texttt{3.1} era. Each tag must be built from a separate Dockerfile,
enabling minor differences in the build instructions to accommodate
changing dependencies. Dockerfiles for past versions (\emph{e.g.}, prior
to \texttt{3.4.2} currently) are intended to remain static over the long
term, while the tag for the current version, \texttt{latest}, and
\texttt{devel} may be tweaked to accommodate new features or
dependencies. Version tags also obey semantics so that omitting the
second or third position of the tag is identical to asking for the most
recent version: i.e. \texttt{rocker/verse:3.3} is the same as
\texttt{rocker/verse:3.3.3}, and \texttt{rocker/verse:3} is (at the time
of writing), \texttt{rocker/verse:3.4.2}. This is accomplished using
post-build hooks in Docker Hub, see examples at
\url{https://github.com/rocker-org/rocker-versioned/} for details.

\textbf{Installation}: In this stack, the desired version of R is always
built directly from source rather than the \texttt{apt} repositories.
Compilers and dependencies are still installed from the stable
\texttt{apt} repositories, and thus lag behind the more recent versions
found in the \texttt{testing} stack. Version tags \texttt{3.3.3} and
older are based on the Debian 8.0 release, code-named \texttt{jessie},
while \texttt{3.4.0} - \texttt{3.4.2}, \texttt{devel}, and
\texttt{latest} are based on Debian 9.0, \texttt{stretch}, (released
2017-06-17, while R was at \texttt{3.4.0}), and thus have access to much
newer versions of common system dependencies and compilers. Dependencies
needed to compile R that are not required at runtime are removed once R
is installed, keeping the base images light-weight for faster download
times. While most system dependencies required by common R packages can
still be installed from the \texttt{apt} repositories, occasionally a
more recent version must be compiled from source (\emph{e.g.}, the Gibbs
Sampling program JAGS \citep{jags}, and the geospatial toolkit GDAL,
must both be compiled from source on \texttt{debian:jessie} images). In
this stack, users should avoid installing R packages using \texttt{apt}
without careful consideration as this will install a second (probably
different) version of R from the Debian repositories, and a dated
version of the R package since any \texttt{r-cran-pkgname} package in
the Debian repositories will depend on \texttt{r-base} in \texttt{apt}
as well.

\textbf{Build schedule}: All images are built automatically from their
corresponding Dockerfiles (found in the GitHub repositories
\texttt{rocker-org/rocker-versioned} and
\texttt{rocker-org/geospatial}). A \texttt{cron} job sends nightly build
triggers to Docker Hub to rebuild the \texttt{latest} and \texttt{devel}
tagged images throughout the stack. To decrease load on the hub, build
triggers for the numeric version tags are sent monthly. Although the
Dockerfiles for older R versions installs an almost-identical software
environment every time, the monthly rebuilding of these images on Docker
Hub ensures they continue to receive Debian security updates from
upstream, and proves the build recipe still executes successfully. Note
that rebuilding images with software from external repositories never
produces a bit-wise identical image, and thus the image identifier hash
will change at each build.

\subsubsection{Images overview}\label{images-overview-1}

In this stack, each image builds on the previous image, rather than all
other images building directly on the base image, as in the
\texttt{testing} stack. Table 2 lists the names and descriptions of the
five images in this stack, along with image size and approximate
download counts from Docker Hub. Sizes reflect (compressed) cumulative
size: a user who has already downloaded the most recent version of
\texttt{r-ver} and then pulls a copy of \texttt{rstudio} image will only
need to download the additional 115 MB in the \texttt{rstudio} layers
and not the full 334 MB listed. This linear design limits flexibility
(no option for \texttt{tidyverse} without \texttt{rstudio}) but
simplifies use and maintenance. While no single environment will be
optimal for everyone, both the packages selected in this stack and the
stack ordering reflect considerable community input and tuning.

The \texttt{rstudio} image includes a lightweight, easy-to-use and
docker-friendly \texttt{init} system, s6 \citep{s6} for running
persistent services, including the
RStudio\textsuperscript{\textregistered} server. This system provides a
convenient way for downstream Dockerfile developers to add additional
persistent services (such as an \texttt{ssh} server) to a single
container, or additional start-up or shutdown scripts that should be run
when a container starts up or shuts down. The \texttt{rstudio} image
uses such a start-up script to configure user settings such as login
password and permissions through environmental variables at run time.

The \texttt{tidyverse} image contains all required and suggested
dependencies of the commonly-used \texttt{tidyverse} and
\texttt{devtools} R packages, including external database libraries
(\emph{e.g.}, MariaDB and PostgreSQL). Users should consult the package
Dockerfiles or \texttt{installed.packages()} list directly for a
complete list of installed packages. The \texttt{verse} library adds
commonly-used dependencies, notably a large but not comprehensive LaTeX
environment and Java development libraries. Previously, the Rocker
project provided the image \texttt{hadleyverse} which was since divided
into \texttt{tidyverse} and \texttt{verse} through community input.

\begin{longtable}[]{@{}lllll@{}}
\caption{Available tags in the \texttt{rocker-versioned}
stack.}\tabularnewline
\toprule
\begin{minipage}[b]{0.09\columnwidth}\raggedright\strut
tag\strut
\end{minipage} & \begin{minipage}[b]{0.12\columnwidth}\raggedright\strut
apt repos\strut
\end{minipage} & \begin{minipage}[b]{0.15\columnwidth}\raggedright\strut
MRAN date\strut
\end{minipage} & \begin{minipage}[b]{0.19\columnwidth}\raggedright\strut
Build frequency\strut
\end{minipage} & \begin{minipage}[b]{0.32\columnwidth}\raggedright\strut
images with tag\strut
\end{minipage}\tabularnewline
\midrule
\endfirsthead
\toprule
\begin{minipage}[b]{0.09\columnwidth}\raggedright\strut
tag\strut
\end{minipage} & \begin{minipage}[b]{0.12\columnwidth}\raggedright\strut
apt repos\strut
\end{minipage} & \begin{minipage}[b]{0.15\columnwidth}\raggedright\strut
MRAN date\strut
\end{minipage} & \begin{minipage}[b]{0.19\columnwidth}\raggedright\strut
Build frequency\strut
\end{minipage} & \begin{minipage}[b]{0.32\columnwidth}\raggedright\strut
images with tag\strut
\end{minipage}\tabularnewline
\midrule
\endhead
\begin{minipage}[t]{0.09\columnwidth}\raggedright\strut
devel\strut
\end{minipage} & \begin{minipage}[t]{0.12\columnwidth}\raggedright\strut
\texttt{stretch}\strut
\end{minipage} & \begin{minipage}[t]{0.15\columnwidth}\raggedright\strut
current date\strut
\end{minipage} & \begin{minipage}[t]{0.19\columnwidth}\raggedright\strut
nightly\strut
\end{minipage} & \begin{minipage}[t]{0.32\columnwidth}\raggedright\strut
\texttt{r-ver}, \texttt{rstudio}, \texttt{tidyverse}, \texttt{verse},
\texttt{geospatial}\strut
\end{minipage}\tabularnewline
\begin{minipage}[t]{0.09\columnwidth}\raggedright\strut
latest\strut
\end{minipage} & \begin{minipage}[t]{0.12\columnwidth}\raggedright\strut
\texttt{stretch}\strut
\end{minipage} & \begin{minipage}[t]{0.15\columnwidth}\raggedright\strut
current date\strut
\end{minipage} & \begin{minipage}[t]{0.19\columnwidth}\raggedright\strut
nightly\strut
\end{minipage} & \begin{minipage}[t]{0.32\columnwidth}\raggedright\strut
\texttt{r-ver}, \texttt{rstudio}, \texttt{tidyverse}, \texttt{verse},
\texttt{geospatial}\strut
\end{minipage}\tabularnewline
\begin{minipage}[t]{0.09\columnwidth}\raggedright\strut
3.4.2\strut
\end{minipage} & \begin{minipage}[t]{0.12\columnwidth}\raggedright\strut
\texttt{stretch}\strut
\end{minipage} & \begin{minipage}[t]{0.15\columnwidth}\raggedright\strut
current date\strut
\end{minipage} & \begin{minipage}[t]{0.19\columnwidth}\raggedright\strut
monthly\strut
\end{minipage} & \begin{minipage}[t]{0.32\columnwidth}\raggedright\strut
\texttt{r-ver}, \texttt{rstudio}, \texttt{tidyverse}, \texttt{verse},
\texttt{geospatial}\strut
\end{minipage}\tabularnewline
\begin{minipage}[t]{0.09\columnwidth}\raggedright\strut
3.4.1\strut
\end{minipage} & \begin{minipage}[t]{0.12\columnwidth}\raggedright\strut
\texttt{stretch}\strut
\end{minipage} & \begin{minipage}[t]{0.15\columnwidth}\raggedright\strut
2017-09-28\strut
\end{minipage} & \begin{minipage}[t]{0.19\columnwidth}\raggedright\strut
monthly\strut
\end{minipage} & \begin{minipage}[t]{0.32\columnwidth}\raggedright\strut
\texttt{r-ver}, \texttt{rstudio}, \texttt{tidyverse}, \texttt{verse},
\texttt{geospatial}\strut
\end{minipage}\tabularnewline
\begin{minipage}[t]{0.09\columnwidth}\raggedright\strut
3.4.0\strut
\end{minipage} & \begin{minipage}[t]{0.12\columnwidth}\raggedright\strut
\texttt{stretch}\strut
\end{minipage} & \begin{minipage}[t]{0.15\columnwidth}\raggedright\strut
2017-06-30\strut
\end{minipage} & \begin{minipage}[t]{0.19\columnwidth}\raggedright\strut
monthly\strut
\end{minipage} & \begin{minipage}[t]{0.32\columnwidth}\raggedright\strut
\texttt{r-ver}, \texttt{rstudio}, \texttt{tidyverse}, \texttt{verse},
\texttt{geospatial}\strut
\end{minipage}\tabularnewline
\begin{minipage}[t]{0.09\columnwidth}\raggedright\strut
3.3.3\strut
\end{minipage} & \begin{minipage}[t]{0.12\columnwidth}\raggedright\strut
\texttt{jessie}\strut
\end{minipage} & \begin{minipage}[t]{0.15\columnwidth}\raggedright\strut
2017-04-21\strut
\end{minipage} & \begin{minipage}[t]{0.19\columnwidth}\raggedright\strut
monthly\strut
\end{minipage} & \begin{minipage}[t]{0.32\columnwidth}\raggedright\strut
\texttt{r-ver}, \texttt{rstudio}, \texttt{tidyverse}, \texttt{verse},
\texttt{geospatial}\strut
\end{minipage}\tabularnewline
\begin{minipage}[t]{0.09\columnwidth}\raggedright\strut
3.3.2\strut
\end{minipage} & \begin{minipage}[t]{0.12\columnwidth}\raggedright\strut
\texttt{jessie}\strut
\end{minipage} & \begin{minipage}[t]{0.15\columnwidth}\raggedright\strut
2017-03-06\strut
\end{minipage} & \begin{minipage}[t]{0.19\columnwidth}\raggedright\strut
monthly\strut
\end{minipage} & \begin{minipage}[t]{0.32\columnwidth}\raggedright\strut
\texttt{r-ver}, \texttt{rstudio}, \texttt{tidyverse}, \texttt{verse},
\texttt{geospatial}\strut
\end{minipage}\tabularnewline
\begin{minipage}[t]{0.09\columnwidth}\raggedright\strut
3.3.1\strut
\end{minipage} & \begin{minipage}[t]{0.12\columnwidth}\raggedright\strut
\texttt{jessie}\strut
\end{minipage} & \begin{minipage}[t]{0.15\columnwidth}\raggedright\strut
2016-10-31\strut
\end{minipage} & \begin{minipage}[t]{0.19\columnwidth}\raggedright\strut
monthly\strut
\end{minipage} & \begin{minipage}[t]{0.32\columnwidth}\raggedright\strut
\texttt{r-ver}, \texttt{rstudio}, \texttt{tidyverse}, \texttt{verse},
\texttt{geospatial}\strut
\end{minipage}\tabularnewline
\begin{minipage}[t]{0.09\columnwidth}\raggedright\strut
3.3.0\strut
\end{minipage} & \begin{minipage}[t]{0.12\columnwidth}\raggedright\strut
\texttt{jessie}\strut
\end{minipage} & \begin{minipage}[t]{0.15\columnwidth}\raggedright\strut
2016-06-21\strut
\end{minipage} & \begin{minipage}[t]{0.19\columnwidth}\raggedright\strut
monthly\strut
\end{minipage} & \begin{minipage}[t]{0.32\columnwidth}\raggedright\strut
\texttt{r-ver}\strut
\end{minipage}\tabularnewline
\begin{minipage}[t]{0.09\columnwidth}\raggedright\strut
3.2.0\strut
\end{minipage} & \begin{minipage}[t]{0.12\columnwidth}\raggedright\strut
\texttt{jessie}\strut
\end{minipage} & \begin{minipage}[t]{0.15\columnwidth}\raggedright\strut
2015-06-18\strut
\end{minipage} & \begin{minipage}[t]{0.19\columnwidth}\raggedright\strut
monthly\strut
\end{minipage} & \begin{minipage}[t]{0.32\columnwidth}\raggedright\strut
\texttt{r-ver}\strut
\end{minipage}\tabularnewline
\begin{minipage}[t]{0.09\columnwidth}\raggedright\strut
3.1.0\strut
\end{minipage} & \begin{minipage}[t]{0.12\columnwidth}\raggedright\strut
\texttt{jessie}\strut
\end{minipage} & \begin{minipage}[t]{0.15\columnwidth}\raggedright\strut
2014-09-17\strut
\end{minipage} & \begin{minipage}[t]{0.19\columnwidth}\raggedright\strut
monthly\strut
\end{minipage} & \begin{minipage}[t]{0.32\columnwidth}\raggedright\strut
\texttt{r-ver}\strut
\end{minipage}\tabularnewline
\bottomrule
\end{longtable}

Several images in the \texttt{rocker-versioned} stack can be customized
on build when built locally (rather than pulling prebuilt images from
Docker Hub) by using the \texttt{-\/-build-arg} option of
\texttt{docker\ build}. In the \texttt{r-ver} image, users can set
\texttt{R\_VERSION} and \texttt{BUILD\_DATE} (MRAN default snapshot). In
the \texttt{rstudio} image users can set \texttt{RSTUDIO\_VERSION}
(otherwise defaults to the most recent version), and the
\texttt{PANDOC\_TEMPLATES\_VERSION} .

This stack also makes use of Docker metadata labels defined by
\url{http://schema-label.org}, indicating image \texttt{license}
(GPL-2.0), \texttt{vcs-url} (GitHub repository), and \texttt{vendor}
(Rocker Project). These metadata can be altered or extended in
downstream images.

\section{Conclusions}\label{conclusions}

Over the past several years, Docker has seen immense adoption across
industry and academia. The Open Container initiative \citep{oci} now
provides an open standard that has further extended this container
approach to research environments through projects such as Singularity
\citep{singularity}, allowing users to deploy containerized environments
such as Rocker on machines where they do not have root access, such as
clusters or private servers. Containerization promises to solve numerous
challenges such as portability and replicability in research computing,
which often relies on complex and heterogeneous software stacks
\citep{Boettiger2015}. Yet implementing such environments in containers
is not a trivial task, and not all implementations provide the same
usability, portability or reproducibility. Here we have detailed the
approach taken by the Rocker project in creating and maintaining these
environments through an open and community-driven process. This
structure of the Rocker project has evolved over three years of
operation while drawing in an ever-widening base of academic
researchers, university instructors and industry users. We believe this
overview will be instructive not only to users and developers interested
in the Rocker project, but as a model for similar efforts around other
environments or domains.

\bibliography{RJreferences}

\begin{thebibliography}{18}
\providecommand{\natexlab}[1]{#1}
\providecommand{\url}[1]{\texttt{#1}}
\expandafter\ifx\csname urlstyle\endcsname\relax
  \providecommand{\doi}[1]{doi: #1}\else
  \providecommand{\doi}{doi: \begingroup \urlstyle{rm}\Url}\fi

\bibitem[Bercot(2017)]{s6}
L.~Bercot.
\newblock \emph{s6: skarnet.org's small and secure supervision software suite},
  2017.
\newblock URL \url{https://skarnet.org/software/s6/}.

\bibitem[Boettiger(2015)]{Boettiger2015}
C.~Boettiger.
\newblock {An introduction to Docker for reproducible research, with examples
  from the R environment}.
\newblock \emph{ACM SIGOPS Operating Systems Review}, 49\penalty0 (1):\penalty0
  71--79, 2015.
\newblock \doi{10.1145/2723872.2723882}.
\newblock URL \url{https://dl.acm.org/citation.cfm?id=2723882
  http://arxiv.org/abs/1410.0846}.

\bibitem[Boettiger and Eddelbuettel(2014)]{edd2014}
C.~Boettiger and D.~Eddelbuettel.
\newblock {Introducing Rocker}: {Docker for R}, 2014.
\newblock URL
  \url{https://ropensci.org/blog/blog/2014/10/23/introducing-rocker}.

\bibitem[Cetinkaya-Rundel and Rundel(2017)]{Mine}
M.~Cetinkaya-Rundel and C.~W. Rundel.
\newblock Infrastructure and tools for teaching computing throughout the
  statistical curriculum.
\newblock \emph{PeerJ Preprints}, 5:\penalty0 e3181v1, Aug. 2017.
\newblock ISSN 2167-9843.
\newblock \doi{10.7287/peerj.preprints.3181v1}.
\newblock URL \url{https://doi.org/10.7287/peerj.preprints.3181v1}.

\bibitem[Clark et~al.(2014)Clark, Culich, Hamlin, and Lovett]{Clark2014}
D.~Clark, A.~Culich, B.~Hamlin, and R.~Lovett.
\newblock {BCE: Berkeley's Common Scientific Compute Environment for Research
  and Education}.
\newblock \emph{Proceedings of the 13th Python in Science Conference (SciPy
  2014)}, pages 1--8, 2014.

\bibitem[Csárdi(2017)]{rhub}
G.~Csárdi.
\newblock \emph{rhub: Connect to 'R-hub', from 'R'}, 2017.
\newblock URL \url{https://github.com/r-hub/rhub}.
\newblock R package version 1.0.1.

\bibitem[{Debian Project}(2017)]{apt_pinning}
{Debian Project}.
\newblock Apt-preferences overview, 2017.
\newblock URL \url{https://wiki.debian.org/AptPreferences}.

\bibitem[Docker(2015)]{what-docker}
Docker.
\newblock What is {Docker}?, 2015.
\newblock URL \url{https://www.docker.com/what-docker}.

\bibitem[Eddelbuettel(2014)]{edd_sanitizers}
D.~Eddelbuettel.
\newblock sanitizers, 2014.
\newblock URL \url{http://dirk.eddelbuettel.com/code/sanitizers.html}.

\bibitem[{LBNL}(2017)]{singularity}
{Lawrence Berkeley National Laboratories}.
\newblock Singularity, 2017.
\newblock URL \url{http://singularity.lbl.gov/}.

\bibitem[Ludaescher et~al.(2017)Ludaescher, Chard, Turk, Stodden, and
  Gaffney]{wholetale}
B.~Ludaescher, K.~Chard, M.~Turk, V.~Stodden, and N.~Gaffney.
\newblock {The Whole Tale}: Merging science and cyberinfrastructure pathways,
  2017.
\newblock URL \url{https://wholetale.org/}.

\bibitem[Plummer(2017)]{jags}
M.~Plummer.
\newblock {JAGS}: {A} program for analysis of {Bayesian} graphical models using
  {Gibbs} sampling, 2017.
\newblock URL \url{http://mcmc-jags.sourceforge.net/}.
\newblock Version 4.3.0.

\bibitem[{Revolution Analytics}(2017)]{MRAN}
{Revolution Analytics}.
\newblock {Microsoft R Application Network}, 2017.
\newblock URL \url{https://mran.microsoft.com}.

\bibitem[Stewart et~al.(2015)Stewart, Turner, Vaughn, Gaffney, Cockerill,
  Foster, Hancock, Merchant, Skidmore, Stanzione, Taylor, and
  Tuecke]{jetstream}
C.~A. Stewart, G.~Turner, M.~Vaughn, N.~I. Gaffney, T.~M. Cockerill, I.~Foster,
  D.~Hancock, N.~Merchant, E.~Skidmore, D.~Stanzione, J.~Taylor, and S.~Tuecke.
\newblock {Jetstream: A self-provisioned, scalable science and engineering
  cloud environment}.
\newblock In \emph{Proceedings of the 2015 XSEDE Conference on Scientific
  Advancements Enabled by Enhanced Cyberinfrastructure - XSEDE '15}, pages
  1--8, New York, New York, USA, 2015. ACM Press.
\newblock ISBN 9781450337205.
\newblock \doi{10.1145/2792745.2792774}.
\newblock URL \url{http://dl.acm.org/citation.cfm?doid=2792745.2792774}.

\bibitem[{The Linux Foundation: Projects}(2017)]{oci}
{The Linux Foundation: Projects}.
\newblock {The Open Container Initiative}, 2017.
\newblock URL \url{https://www.opencontainers.org/}.

\bibitem[{UC Berkeley}(2017)]{data8}
{UC Berkeley}.
\newblock {Curriculum Overview} | {Division of Data Sciences}, 2017.
\newblock URL \url{http://data.berkeley.edu/education/curriculum-overview}.

\bibitem[Ushey et~al.(2016)Ushey, McPherson, Cheng, Atkins, and
  Allaire]{packrat}
K.~Ushey, J.~McPherson, J.~Cheng, A.~Atkins, and J.~Allaire.
\newblock \emph{packrat: A Dependency Management System for Projects and their
  R Package Dependencies}, 2016.
\newblock URL \url{https://CRAN.R-project.org/package=packrat}.
\newblock R package version 0.4.8-1.

\bibitem[Wickham(2017)]{tidyverse}
H.~Wickham.
\newblock \emph{tidyverse: Easily Install and Load 'Tidyverse' Packages}, 2017.
\newblock URL \url{https://CRAN.R-project.org/package=tidyverse}.
\newblock R package version 1.1.1.

\end{thebibliography}

\address{%
Carl Boettiger\\
UC Berkeley\\
ESPM Department, University of California,\\ 130 Mulford Hall Berkeley, CA 94720-3114, USA\\
}
\href{mailto:cboettig@berkeley.edu}{\nolinkurl{cboettig@berkeley.edu}}

\address{%
Dirk Eddelbuettel\\
Debian and R Projects; Ketchum Trading\\
Chicago, IL, USA\\
}
\href{mailto:edd@debian.org}{\nolinkurl{edd@debian.org}}

\end{article}

\end{document}